\def\Title{Perfect Computational Equivalence between\\ 
Quantum Turing Machines and\\ 
Finitely Generated Uniform Quantum Circuit Families}
\def\Author{Harumichi Nishimura$^{1}$ and Masanao Ozawa$^{2}$}
\documentstyle[12pt]{article}

  \newcommand{\beq}{\begin{equation}}
  \newcommand{\eeq}{\end{equation}}
  \newcommand{\beql}[1]{\begin{equation}\label{eq:#1}}

  \newcommand{\beqa}{\begin{eqnarray}}
  \newcommand{\eeqa}{\end{eqnarray}}
  \newcommand{\beqas}{\begin{eqnarray*}}
  \newcommand{\eeqas}{\end{eqnarray*}}

  \newtheorem{Theorem}{Theorem}[section]
  \newtheorem{Proposition}[Theorem]{Proposition}
  \newtheorem{Lemma}[Theorem]{Lemma}
  \newtheorem{Corollary}[Theorem]{Corollary}

  \newenvironment{Proof}{\begin{trivlist}
    \item[\hskip \labelsep {\em \indent Proof.}]}{\qed\end{trivlist}}
  \newcommand{\qed}{{\em QED}}

  \newcommand{\de}{\delta}                                             %

  \newcommand{\si}{\sigma}                                             %
  \newcommand{\ta}{\tau}                                               %
  \newcommand{\Si}{\Sigma}

  \newcommand{\beqan}{\begin{eqnarray*}}
  \newcommand{\beqar}[1]{\begin{equation}\label{#1}\begin{array}{l}}

  \newcommand{\eeqar}{\end{array}\end{equation}}

\newcommand{\ignore}[1]{}

  \title{\bf \Title}
\author{\rm \Author \\ \\
  \small\rm ${}^{1}$ School of Science, Osaka Prefecture University, Sakai 599-8531, Japan\\ 
 \small\tt hnishimura@mi.s.osakafu-u.ac.jp\\
  \small\rm ${}^{2}$ Graduate School of Information Science, Nagoya University, Nagoya 464-8601, Japan\\
  \small\tt ozawa@is.nagoya-u.ac.jp
}

  \date{}

\begin{document}

\maketitle

\begin{abstract}
In order to establish the computational equivalence between quantum Turing machines (QTMs) and 
quantum circuit families (QCFs) using Yao's quantum circuit simulation of QTMs, 
we previously introduced the class of uniform QCFs based on an infinite set of elementary gates, 
which has been shown to be computationally equivalent to the polynomial-time QTMs 
(with appropriate restriction 
of amplitudes) up to bounded error simulation. This result implies that the complexity class {\bf BQP} introduced 
by Bernstein and Vazirani for QTMs equals its counterpart for uniform QCFs. 
However, the complexity classes {\bf ZQP} and {\bf EQP} for QTMs do not appear to equal 
their counterparts for uniform QCFs.    
In this paper, we introduce a subclass of uniform QCFs, the {\em finitely generated uniform QCFs}, 
based on finite number of elementary gates and show that the class of finitely generated uniform QCFs 
is {\em perfectly equivalent} to the class of polynomial-time QTMs; they can exactly simulate each other. 
This naturally implies that {\bf BQP} as well as {\bf ZQP} and {\bf EQP} equal 
the corresponding complexity classes of the finitely generated uniform QCFs.  
\end{abstract}



\section{Introduction}
In computational complexity theory, 
Turing machines and Boolean circuits are commonly used as mathematical models of
computation. A Turing machine
has tapes of infinite-length to treat the inputs of any length. 
On the other hand, a Boolean circuit can only process the inputs of a fixed length 
since its size is finite. To represent an algorithm carried out on inputs of any length, 
we need to consider a family of Boolean circuits whose $n$-th circuit handles 
the inputs of length $n$.  In addition, to keep the computational power
consistent with the Church-Turing thesis, we need the notion of uniformity 
of circuit families, first proposed by Borodin \cite{Bor77}.
Roughly speaking, the uniformity of a circuit family 
requires that there exists a computationally simple rule for constructing all circuits 
in the family. It is well-known that the polynomial-time Turing machine and 
the polynomial-size uniform circuit family can exactly simulate each 
other (see, for instance, Theorem 11.5 in \cite[p.~269]{Pap94}). 
We say that two classes $A$ and $B$ of computational models, such as the
class of polynomial-time Turing machines and the class of polynomial-size 
uniform circuit families, are {\em perfectly equivalent} if any element of $A$ 
can be exactly simulated by an element of $B$ in a sense appropriately defined thereon, 
and vise versa. Thus, the class of polynomial-time Turing machines is {perfectly equivalent} 
to the class of polynomial-size uniform circuit families. 

In the mid-1980s, Deutsch proposed a new parallel computing paradigm, 
{\em quantum computing}, which utilizes the superposition principle of quantum mechanics. 
As mathematical models of quantum computing, he introduced the quantum versions of 
Turing machines and Boolean circuits, called quantum Turing machines (QTMs) 
and quantum circuits \cite{Deu85,Deu89}. 
In 1993, Bernstein and Vazirani \cite{BV93} and Yao \cite{Yao93} 
reformulated Deutsch's QTM model and quantum circuit model in the form suitable 
for computational complexity theory. 
Bernstein and Vazirani introduced the complexity classes 
{\bf BQP}, {\bf ZQP}, and {\bf EQP} that represent the bounded-error, zero-error, 
and exact quantum algorithms to set the grounds for quantum complexity theory. 
As the first relation between QTMs and quantum circuits, Yao showed that, 
for any QTM $M$ and $T>0$, there is a quantum circuit of size $O(T^2)$ that simulates $M$ for $T$ steps. 
This result was often mentioned to directly mean that QTMs and quantum circuit 
families (QCFs) are {\em equivalent} by a simple analogy with the equivalence between 
classical Turing machines and classical circuit families. 
After a few years, the notion of uniformity of QCFs was shortly mentioned in \cite{EJ96,Sho97}. 
In particular, Shor \cite{Sho97} pointed out (after his private communication with R. Solovay) 
that uniform QCFs should satisfy a requirement, which is absent in the standard uniformity requirement,
that the entries of elementary gates be polynomial-time computable. 
In tandem, Adleman and collaborators \cite{ADH97} showed that 
a polynomial-time QTM with amplitudes from the whole complex number field 
can compute even a non-recursive function. In the journal version \cite{BV97} of \cite{BV93}, 
Bernstein and Vazirani defined the complexity classes 
{\bf BQP}, {\bf ZQP}, and {\bf EQP} based on the polynomial-time QTM with amplitudes 
from the polynomial-time computable numbers.

In our previous investigation \cite{NO02}, we instituted the complexity theory of uniform QCFs . 
For this purpose, we rigorously introduced the notion of uniformity of QCFs 
and based on that we defined the complexity classes, what we called {\bf BUPQC}, {\bf ZUPQC}, and {\bf EUPQC}, 
which correspond to {\bf BQP}, {\bf ZQP}, and {\bf EQP}, respectively. 
Using this formulation and Yao's quantum circuit construction for simulation of QTMs, 
we showed the following results on the computational equivalence between QTMs and uniform QCFs: 
(i) ${\bf BUPQC}={\bf BQP}$ \cite{NO02} (QTMs and uniform QCFs are computationally equivalent 
in the bounded-error setting). 
(ii) ${\bf ZQP}\subseteq{\bf ZUPQC}$ and ${\bf EQP}\subseteq{\bf EUPQC}$ while 
the simulation does not work to show the converse \cite{NO02,NO05} 
(the computational equivalence between the two models are open in the zero-error and exact setting). 
Thus, the following question still remained: In the zero-error and exact setting, 
what restriction for the uniform QCF guarantees the computational equivalence between the two models? 
Kitaev and Watrous \cite{KW00} introduced the notion of {\em uniformly generated QCFs} 
based on Shor basis \cite{Sho96}. 
The class of uniformly generated QCFs is computationally equivalent 
to the class of our uniform QCFs up to bounded error simulation, 
and it has been used for the study of quantum interactive proof systems. 
However, this notion does not provide a solution to question (ii) above.   

In this paper, we give a complete answer to the above question. 
To this end, we introduce the notion of {\em finitely generated uniform QCFs} based on finite sets 
of elementary gates, which is a subclass of the uniform QCFs. 
In \cite{NO02}, we briefly mentioned the notion of semi-uniform QCFs, 
which is also based on finite sets of elementary gates. 
Finitely generated uniform QCFs are regarded as those semi-uniform QCFs whose amplitudes 
are taken from polynomial-time computable numbers. 
We show that the class of finitely generated uniform QCFs is {\em perfectly equivalent} 
to the class of polynomial-time QTMs with amplitudes from polynomial-time computable numbers. 
This implies that these two models can simulate each other without error, 
and hence not only {\bf BQP} but also {\bf ZQP} and {\bf EQP} coincide with the corresponding classes 
defined through finitely generated uniform QCFs. 
The proof requires more minute arguments than that of the computational equivalence 
up to bounded error simulation. For the proof, we revisit properties of polynomial-time computable numbers, 
which was implicitly used in \cite{NO02}, and Yao's quantum circuit construction.
We combine them with the exact decomposition of unitary matrices.

\section{Finitely Generated QCFs}
In this section, we give a formal definition of finitely generated QCFs. 
For the definition of QCFs, see \cite{NO02}. 
For convenience, we usually identify quantum gates or circuits with the unitary matrices representing them in the computational basis.  
For other fundamental notions of quantum computation, we refer to Gruska \cite{Gru99} and to Nielsen and Chuang \cite{NC00}. 
For classical complexity theory, we refer, for example, to Papadimitriou \cite{Pap94}.  
In what follows, let ${\bf N}$ and ${\bf C}$ denote the sets of natural numbers and complex numbers, 
respectively. Let ${\rm P}{\bf C}$ denote the set of polynomial-time computable complex numbers \cite{KF82}. 
(Informally speaking, a complex number $r$ is polynomial-time computable if its real and imaginary parts can be approximated with accuracy of $1/2^n$ in time polynomial in $n$. See Ko \cite{Ko91} for its formal definition.) 
For two integers $m$ and $n$ with $m<n$, let $[m,n]_{\bf Z}$ denote $\{m,m+1,\ldots,n\}$. 

The precise formulation of the uniformity of QCFs was introduced in our previous work \cite{NO02}. 
Let ${\cal G}_u$ be the set of quantum gates such that 
$$
{\cal G}_u
=\{\Lambda_1(N),R(\theta),P(\theta')|\ \theta,\theta'\in{\rm P}{\bf C}\cap[0,2\pi) \},
$$  
where $\Lambda_1(N)$ is a controlled-not gate, $R(\theta)$ is a rotation gate by angle $\theta$, and $P(\theta')$ is a phase shift gate by angle $\theta'$.  
All the gates in ${\cal G}_u$ can be encoded by binary strings, using the {\em codes} of polynomial-time computable numbers. 
Here, the code of a polynomial-time computable number $r$ is an appropriate encoding of a deterministic Turing machine (DTM) that approximates $r$ within $2^{-n}$ in time polynomial in $n$. 
We can then give the code ${\rm Code}(C)$ of a quantum circuit $C$ based on ${\cal G}_u$. 
We say that a QCF ${\cal C}=\{C_n\}$  based on ${\cal G}_u$ is {\em polynomial-size uniform}, or {\em uniform} for short, if the function $1^n\mapsto {\rm Code}(C_n)$ is computable by a DTM in time polynomial in $n$. Uniform QCFs are very suitable to represent important quantum algorithms such as the quantum Fourier transform and the amplitude amplification.  
However, uniform QCFs cannot be exactly simulated by any QTM \cite{NO02} (even with zero-error \cite{NO05}.) 
Thus, it is questionable whether the class of languages recognized by uniform QCFs with certainty (resp.\ with zero-error) coincides with the class {\bf EQP} (resp.\ {\bf ZQP}) of polynomial-time QTMs.   

By analogy with the uniformity of classical circuits, we may imagine a ``uniform'' QCF based on a fixed finite set of elementary gates. 
For example, Kitaev and Watrous \cite{KW00} introduced {polynomial-time uniformly generated QCFs} to define verifiers of quantum interactive proof systems: A QCF ${\cal C}=\{C_x\}$\footnote{Their family is somewhat nonstandard in the sense that the parameter of the family is a binary string that represents an input, not an integer that represents the length of the input. However, as they mentioned in \cite{KW00}, this does not change the computational power for some fixed set of elementary gates. 
(Also, see Section \ref{Discussions}.)} is {\em polynomial-time uniformly generated} if there exists a deterministic procedure that, on input $x$, outputs a description of $C_x$ and runs in time polynomial in $|x|$. 
Although they used Shor's basis \cite{Sho96} as their set of elementary gates, in the bounded-error setting, any other universal set of elementary gates is available without changing the computational power of polynomial-time uniformly generated QCFs. 
However, for exact (or zero-error) algorithms, fixing a set of elementary gates may seriously reduce the computational power of ``uniform'' QCFs. 
For instance, the class of languages recognized with bounded-error by polynomial-time uniformly generated QCFs 
based on a universal set ${\cal G}_{4/5}=\{ \Lambda_1(N),R(\theta),P(\theta)|\ \cos\theta=4/5\}$ coincides with the class {\bf BQP} \cite{ADH97}, while the class of languages recognized with certainty by polynomial-time uniformly generated QCFs based on ${\cal G}_{4/5}$ coincides with the class {\bf P} \cite{Nis03}, instead of the class {\bf EQP}.  

We now introduced a class of QCFs, called {\em finitely generated QCFs}, an intermediate class between uniform QCFs and polynomial-time uniformly generated QCFs. 
A uniform QCF ${\cal C}$ is said to be {\em finitely generated} if there is a finite subset ${\cal G}$ of ${\cal G}_u$  such that ${\cal C}$ is based on ${\cal G}$. 
By definition, any polynomial-time uniformly generated QCF is finitely generated, and any finitely generated uniform QCF is uniform. Finitely generated uniform QCFs have two nice properties: (i) The finitely generated uniform QCF is based on finite sets of elementary gates, different from the uniform QCF; and (ii) The definition of the finitely generated uniform QCF is independent of the choice of universal sets, different from the polynomial-time uniformly generated QCF. 
In \cite{NO02}, we provided a similar notion, called semi-uniform QCFs. 
A finitely generated uniform QCF can be regarded as a semi-uniform QCF such that all the components representing matrices of elementary gates are restricted to ${\rm P}{\bf C}$. 
An analogous concept to finitely generated uniform QCFs was also mentioned by Green, Homer, Moore, and Pollett \cite{GHMP02} in their study of shallow quantum circuits.

\section{Proof of the Perfect Equivalence}
In this section, we establish the perfect equivalence between the finitely generated uniform QCF and the polynomial-time QTM with amplitudes from ${\rm P}{\bf C}$. 
See \cite{BV97,NO02,ON00,Yam99} for the definition of QTMs.   

First, we recall two elementary properties on polynomial-time computable numbers given by Ko and Friedman \cite{KF82}.  
Loosely speaking, a real function $f$ is said to be polynomial-time computable if for any real $x$, the value $f(x)$ can be approximated with accuracy of $1/2^n$ in time polynomial in $n$ using $x$ as an ``oracle'' to obtain any required bits 
of $x$. (See \cite{Ko91} for its formal definition.) 

\begin{Theorem}\label{pc-property} 
(i) \sloppy All roots of an analytic, polynomial-time computable function 
are polynomial-time computable.  

(ii) ${\rm P}{\bf C}$ is an algebraically closed field.  
\end{Theorem} 

Second, we provide a standard exact decomposition of finite dimensional unitary matrices \cite{BBC$^+$95} 
with an argument on polynomial-time computable entries of matrices.\footnote{Lately, more efficient decomposition is shown (e.g., \cite{MV05}) but they do not reduce the number of gates exponentially. Thus, for simplicity of our argument, we use the decomposition in \cite{BBC$^+$95}.} To this end, we use the terminology of the approximate decomposition algorithm of unitary matrices given in \cite{BV97}. 
Although this exact decomposition is not algorithmic, it is sufficient for our purpose. 
Let ${\bf e}_j$ be the $m$-dimensional unit column vector whose $j$-th component is 1. 
We denote by ${\rm Near}_m(j,\theta)$ an $m$-dimensional unitary matrix satisfying 
$$
{\rm Near}_m(j,\theta){\bf e}_k=
\left\{
\begin{array}{ll}
(e^{\imath\theta}) {\bf e}_k & \mbox{if}\ k=j,\\
               {\bf e}_k         & \mbox{otherwise}.
\end{array}
\right.
$$
We denote by ${\rm Near}_m(i,j,\theta)$ an $m$-dimensional unitary matrix satisfying
$$
{\rm Near}_m(i,j,\theta){\bf e}_k=
\left\{
\begin{array}{ll}
(\cos\theta) {\bf e}_i - (\sin\theta) {\bf e}_j & \mbox{if}\ k=i,\\
(\sin\theta) {\bf e}_i + (\cos\theta) {\bf e}_j & \mbox{if}\ k=j,\\
{\bf e}_k                             & \mbox{otherwise}.
\end{array}
\right.
$$  
These two types of matrices are called {\em near-trivial} \cite{BV97}. 
We show that any unitary matrix whose components are in {\rm P}{\bf C} can be decomposed into near-trivial matrices whose components are also in {\rm P}{\bf C}. 

\begin{Lemma}\label{exact-decomposition-1} 
(i) Any $N$-dimensional unitary matrix $U$ can be represented by the product $U_m\cdots U_1$ of $m=O(N^2)$ near-trivial matrices $U_1,\ldots,U_m$. 

(ii) Moreover, if all the entries in $U$ are in ${\rm P}{\bf C}$, each $U_j$ has the entries in ${\rm P}{\bf C}$.
\end{Lemma}

\begin{Proof} (i) Let $U$ be an $N$-dimensional unitary matrix. 
We can show that there is a product $A$ of $2N-1$ near-trivial matrices such that $AU^{(1)}={\bf e}_1$, where $U^{(1)}$ denotes the first column vector of $U$. 
Actually, we show that for any $N$-dimensional unit column vector ${\bf v}$, there is a product $A$ of $2N-1$ near-trivial matrices such that $A{\bf v}={\bf e}_1$. 
Let $v_i$ denote the $i$-th coordinate of ${\bf v}$. 
First, we use $N$ near-trivial matrices to map ${\bf v}$ into the $N$-dimensional real space. Let $P_i={\rm Near}_m(i,\phi_i)$ where
$$
\phi_j =\left\{ 
\begin{array}{ll} 
2\pi-\cos^{-1}\left(\frac{{\rm Re}(v_j)}{|v_j|} \right)
&\ \mbox{if}\ {\rm Im}(v_j)>0,\\ 
\cos^{-1}\left(\frac{{\rm Re}(v_j)}{|v_j|}\right)
&\ \mbox{if}\ {\rm Im}(v_j)<0,\\ 
0&\ \mbox{if}\ v_j=0.
\end{array}
\right. 
$$
Then, $P_1\cdots P_N {\bf v}$ is the vector with $i$-th coordinate $|v_i|$. 
\sloppy
Second, we use $N-1$ near-trivial matrices to move all of the weight of the vector into dimension 1. 
Let $R_i={\rm Near}_m(i,i+1,\theta_i)$, where $\theta_i=\tan^{-1}(\sqrt{\sum_{j=i+1}^N|v_j|^2}/|v_i|)$. 
Then, we have $R_1\cdots R_{N-1}P_1\cdots P_N {\bf v}={\bf e}_1$. 

Now we have $AU={\rm diag}(1,B)$, where $B$ is an $(N-1)$-dimensional unitary matrix. 
Here, ${\rm diag}(A_1,\ldots,A_n)$ denotes the block-diagonal square matrix composed of square matrices $A_1,\ldots,A_n$ along the diagonal and 0's everywhere else. 
By induction, we can verify that there is the product $C$ of $O(N^2)$ near-trivial matrices 
satisfying $CU=I$. Thus, $U$ can be represented by the product of $O(N^2)$ near-trivial matrices. 

(ii) In the case where all the entries in $U$ are in ${\rm P}{\bf C}$, 
we should note that the above decomposition uses only four arithmetic operations, 
the function $x\mapsto \sqrt{x}$, the trigonometric functions and their inverse functions 
for each real parts and imaginary parts. 
Therefore, each near-trivial matrix composing $U$ has the entries in ${\rm P}{\bf C}$ by Theorem \ref{pc-property}. 
\end{Proof}

As seen in \cite{BBC$^+$95}, 
any $2^n$-dimensional near-trivial unitary matrix $U$ can be decomposed into 
$O(n^3)$ quantum gates in ${\cal G}_u$ as follows: (i) Using an idea from the grey code, 
$U$ can be decomposed into $O(n)$ controlled$^{(n-1)}$-phase shifts or rotation gates; (ii) Any controlled$^{(n-1)}$-phase shift (or rotation gate) can be decomposed into $O(n^2)$ quantum gates in ${\cal G}_u$ (Corollary 7.6 in \cite{BBC$^+$95}). 
Moreover, by carefully checking the argument leading to obtain Corollary 7.6 in \cite{BBC$^+$95}, 
we can see that those $O(n^3)$ quantum gates have their entries in ${\rm P}{\bf C}$ 
provided that all the entries in $U$ are in ${\rm P}{\bf C}$. 
Thus, we obtain the following exact decomposition of unitary matrices.  

\begin{Proposition}\label{exact-decomposition} 
(i) Any $n$-qubit gate $U$ can be decomposed into $O(2^{2n}n^3)$ quantum gates in ${\cal G}_u$. 

(ii) Moreover, if all the entries in $U$ are in ${\rm P}{\bf C}$, each of the quantum gates that decomposes $U$ has its entries in ${\rm P}{\bf C}$.
\end{Proposition}

Third, we provide the exact simulation of QTMs by finitely generated uniform QCFs, combining Yao's idea for the simulation of a QTM by a quantum circuit \cite{Yao93} and Proposition \ref{exact-decomposition}. 
We briefly recall the definition of the simulation of QTMs by finitely generated uniform QCFs. 
See \cite{NO02} for the detail. A quantum circuit $C$ is said to {\em exactly $t$-simulate} a QTM $M$ 
if the following two probability distributions ${\cal D}_x$ and ${\cal D}'_x$ are equal for any string $x$: 
(i) the probability distribution ${\cal D}_x$ of the outcomes of the simultaneous measurement of the tape cells 
from cell $-t$ to cell $t$ after $t$ steps of $M$ for input state $|q_0,{\rm tape}[x],0\rangle$, where ${\rm tape}[x]$ represents the tape configuration such that $x$ is written from cell 0 to cell $|x|-1$; and (ii) the probability distribution ${\cal D}'_x$ of the string obtained by decoding the output of $C$ for the input 
of the binary string obtained by encoding $x$. A finitely generated uniform QCF $\{C_n\}$ 
is said to {\em exactly simulate} a polynomial-time QTM $M$ if $C_n$ $t(n)$-simulates $M$, 
where a polynomial $t(n)$ is the computation time of $M$ on input of length $n$. 

\begin{Theorem}\label{exact-equiv} 
Let $M=(Q,\Sigma,\delta)$ be a polynomial-time QTM with amplitudes from ${\rm P}{\bf C}$. 
Then, there is a finitely generated uniform QCF $\{C_n\}$ that exactly simulates $M$.  
\end{Theorem}

\begin{Proof} The basic line of this proof is in accordance with the proof in \cite[Theorem 4.3]{NO02}, which is based on Yao's construction. For our exact simulation, we argue the decomposition of Yao's construction into elementary gates in detail. We refer to \cite[Section 4]{NO02} for the terminology of quantum circuits. 

Let $t(n)$ be the computation time of $M$ on input of length $n$. First, we fix $n$ and construct a quantum circuit $C_{\cal G}$ which $t$-simulates $M$. Henceforth, let $t=t(n)$ for simplicity. The quantum gate determined by $C_{\cal G}$ consists of $l_0+(2t+1)l$ wires (i.e., qubits), where $l_0=\lceil \log |Q|\rceil$ and $l=2+\lceil \log |\Sigma| \rceil$. Its wires are indexed in order. We divide their wires into a part consisting of the first $l_0$ wires and $2t+1$ parts. Each of the $2t+1$ parts consists of $l$ wires. The part consisting of the first $l_0$ wires represents the processor configuration of $M$. This set of wires is called cell ``P'' of $C_{\cal G}$. 
The state of the cell P of $C_{\cal G}$ is represented by a unit vector in the Hilbert space spanned by the computational basis $\{|q\rangle\}$, where $q\in\{0,1\}^{l_0}$. 
For $j\in[0,2t]_{\bf Z}$, the wires with indices $l_0+jl+1,\ldots,l_0+jl+l$ represent the symbol in the $(j-t)$-th cell of $M$ and whether the head scans this cell or not. This set of wires is called cell $j-t$ of $C_{\cal G}$. 
For $i\in[-t,t]_{\bf Z}$, the state of the cell $i$ of $C_{\cal G}$ is represented by a unit vector in the Hilbert space spanned by the computational basis $\{ |\si_is_i\rangle \}$, where $\si_i\in\{0,1\}^{\lceil \log |\Si| \rceil}$ and $s_i\in\{0,1\}^2$. 

The circuit $C_{\cal G}$ consists of two types of quantum gates $G_1$ and $G_2$. 
The quantum gate $G_1$ is used for simulating one step of $M$ 
when its gate are connected into the cell in which the head of $M$ exists. 
The quantum gate $G_2$ is used for resetting some wires to simulate the next step. 
In what follows, $p,q,\ldots$ denote binary strings representing elements of $Q$, the symbols $\si,\ta,\ldots$ denote binary strings representing elements of $\Si$, and $s=\bar{0},\bar{1},\bar{2}$ denote 00,01,10, respectively. 
We denote by $|q;\si_1s_1;\si_2s_2;\cdots;\si_ks_k\rangle$ the computational basis state $|q\si_1s_1\si_2s_2\cdots \si_ks_k\rangle$ on the wires corresponding to all the numbers in $[1,l_0+kl]_{\bf Z}$. 
Let $G_1$ be an $(l_0+3l)$-qubit gate satisfying the following conditions (i) and (ii):

(i) $G_1|w_{p,\si_1,\si,\si_3}\rangle=|v_{p,\si_1,\si,\si_3}\rangle$, where 
\begin{eqnarray*}
 |w_{p,\si_1,\si,\si_3}\rangle
 &=& |p; \si_1\bar{0};\si \bar{1};\si_3\bar{0}\rangle,\ \ \mbox{and}\\
 |v_{p,\si_1,\si,\si_3}\rangle
 &=&\sum_{q,\ta}\de(p,\si,q,\ta,-1) |q; \si_1\bar{2};\ta \bar{0};\si_3\bar{0}\rangle \\
& &  \mbox{ } +\sum_{q,\ta}\de(p,\si,q,\ta,0)  |q; \si_1\bar{0};\ta \bar{2};\si_3\bar{0}\rangle\\
 & &\mbox{ }
   +\sum_{q,\ta}\de(p,\si,q,\ta,1)  |q; \si_1\bar{0};\ta \bar{0};\si_3\bar{2}\rangle
\end{eqnarray*}
for any $(p,\si_1,\si,\si_3)\in Q\times\Si^3$; each summation $\sum_{q,\ta}$ is taken over all $(q,\ta)\in Q\times \Si$.

(ii) $G_1|h\rangle=|h\rangle$ for each vector $|h\rangle$ in the subspace $H$ of ${\bf C}^{2^{l_0+3l}}$ 
spanned by three types of vectors:

(1) $|q;\si_1s_1;\si_2s_2;\si_3s_3\rangle$, where $s_2\neq\bar{1}$ and none of $s_1,s_2,s_3$ equals $\bar{2}$;

\medskip
(2) $|u^1_{p,\si,\si_2,\si_3}\rangle = \sum_{q,\ta}\de(p,\si,q,\ta,0)|q; \ta\bar{2};\si_2\bar{0};\si_3\bar{0}\rangle 
$\\
$\hspace{\fill}+\sum_{q,\ta}\de(p,\si,q,\ta,1) |q; \ta\bar{0};\si_2\bar{2};\si_3\bar{0}\rangle;$

\medskip
(3) $|u^2_{p,\si,\ta,\si_1,\si_2,\si_3}\rangle = \sum_{q\in Q}\de(p,\si,q,\ta,1)
|q; \si_1\bar{2};\si_2\bar{0};\si_3\bar{0}\rangle$.

\medskip
\ignore{
Let $W=\{ |w_{p,\si,\si_1,\si_3}\rangle |\ (p,\si,\si_1,\si_3)\in Q\times\Si^3\}^{\bot\bot}$ and $V=\{ |v_{p,\si,\si_1,\si_3}\rangle |\ (p,\si,\si_1,\si_3)\in Q\times\Si^3\}^{\bot\bot}$, where $S^{\bot}$ denotes the orthogonal complement of a set $S$ so that $S^{\bot\bot}$ denotes the subspace generated by $S$. 
By the characterization of quantum transition function $\delta$ of $M$ \cite{ON00}, we can check that $W$, $V$ and $H$ are all orthogonal one another and that $\{ |v_{p,\si,\si_1,\si_3}\rangle \}$ is an orthonormal system of $V$. 
Thus, there exists a quantum gate $G_1$ satisfying the above condition. 
}
\noindent Let $G_2$ be an $(l_0+(2t+1)l)$-qubit gate that does nothing except for mapping all $s_i=\bar{2}$'s to $s_i=\bar{1}$'s and vice versa. 
\ignore{
Henceforth, given any $m\in[1,2t+1]_{\bf Z}$, we say that an $(l_0+ml)$-qubit gate $G$ is connected with cells $i_1,\ldots,i_m$, where $i_1<\cdots <i_m$ if each $j_0$-th pin of $G$, for $j_0\in[1,l_0]_{\bf Z}$, and each $(l_0+jl-l+k)$-th pin of $G$, for $j\in[1,m]_{\bf Z}$ and $k\in[1,l]_{\bf Z}$, are respectively connected with the wires of numbers $j_0$ and $l_0+(i_j+t)l+k$.  
}
Let $C_{\cal G}$ be the quantum circuit based on ${\cal G}=\{G_1,G_2\}$ constructed as follows. 
First, $2t-1$ $G_1$'s are connected in such a way that, for $j\in[1,2t-1]_{\bf Z}$, the $j$-th $G_1$ is connected with cells $j-t-1,j-t$ and $j-t+1$.  
The $(l_0+(2t+1)l)$-qubit circuit constructed from these $G_1$'s is called $C_1$.  
Lastly, $G_2$ is connected with cells $-t,-t+1,\ldots,t$.  
The $(l_0+(2t+1)l)$-qubit circuit constructed from this $G_2$ is called $C_2$. 
Let $C_{\cal G}$ be $(C_2\circ C_1)^t$, i.e., the $t$ concatenations of the circuit $C_2\circ C_1$ obtained by connecting $C_2$ into $C_1$. 
The quantum circuit $C_2\circ C_1$ is illustrated in Figure 1. 
By the definitions of $G_1$ and $G_2$, it can be verified that $C_2\circ C_1$ carries out the operation corresponding to one step of $M$. (See \cite{NO02} for the verification.) 
Hence, $C_{\cal G}$ simulates $t$ steps of $M$ exactly. 

The $(l_0+3l)$-qubit gate $G_1$ can be exactly decomposed by a finite number of one-qubit gates $G_{11},\ldots,G_{1\alpha}$ and $\Lambda_1(N)$ from Proposition \ref{exact-decomposition}(i). 
Note that all the components of $G_{11},\ldots,G_{1\alpha},\Lambda_1(N)$ are in ${\rm P}{\bf C}$ by Proposition \ref{exact-decomposition}(ii) since the QTM $M$ has transition amplitudes in ${\rm P}{\bf C}$.  
By definition, the quantum gate $G_2$ can be implemented by swapping two qubits of each $s_i$, which can be implemented by the concatenation of three controlled-not gates. 
Thus, there are an $(l_0+3l)$-qubit quantum circuit $C_{u,1}$ of size $O(1)$ and an $(l_0+(2t+1)l)$-bit quantum circuit $C_{u,2}$ of size $3(2t+1)$ based on ${\cal G}_u$ such that the quantum gates determined by them are $G_1$ and $G_2$, respectively. 
Now, let $C_a$ be an $(l_0+(2t+1)l)$-qubit quantum circuit obtained by decomposing each $G_1$ in $C_1$ into $O(1)$ gates using a subset ${\cal G}_M=\{ G_{11},\ldots,G_{1\alpha},\Lambda_1(N) \}$ of ${\cal G}_u$. 
Since the size of $C_a$ is $O(2t+1)$, $C=(C_{u,2}\circ C_a)^t$ is a quantum circuit based on ${\cal G}_M$ of size $O(t^2)$ that exactly $t$-simulates $M$. The codes of $C_{u,2}$ and $C_a$ can be computed by a DTM in time polynomial in $n$ from their constructions (note that the codes of polynomial-time computable numbers used in $C_{u,2}$ and $C_a$ can be stored 
in the processor of the DTM since they are finite). 
Moreover, the code of $C$ is also computed in time polynomial in $n$ since $C$ is simply the concatenation of $t(n)$ circuits $C_{u,2}\circ C_a$. 
Noting that the set ${\cal G}_M$ of elementary gates is fixed with respect to $n$, we can verify that $M$ is exactly simulated by a finitely generated uniform QCF $\{C_n\}$, where $C_n$ is the above quantum circuit that exactly $t(n)$-simulates $M$. 
This completes the proof.  
\end{Proof}

\begin{figure}[bt]
\begin{center}\unitlength 2.5mm
\begin{picture}(47,22)
\put(8,3){\line(0,1){2}}\put(9,3){\line(0,1){2}}
\put(12,3){\line(0,1){2}}\put(13,3){\line(0,1){2}}
\put(16,3){\line(0,1){2}}\put(17,3){\line(0,1){2}}
\put(20,3){\line(0,1){4}}\put(21,3){\line(0,1){4}}
\put(41,3){\line(0,1){12}}\put(42,3){\line(0,1){12}}
\put(7,5){\framebox(11,1)}

\put(8,6){\line(0,1){11}}\put(9,6){\line(0,1){11}}
\put(12,6){\line(0,1){1}}\put(13,6){\line(0,1){1}}
\put(16,6){\line(0,1){1}}\put(17,6){\line(0,1){1}}
\put(11,7){\framebox(11,1)}
\put(12,8){\line(0,1){9}}\put(13,8){\line(0,1){9}}
\put(16,8){\line(0,1){1}}\put(17,8){\line(0,1){1}}
\put(20,8){\line(0,1){1}}\put(21,8){\line(0,1){1}}
\put(33,16){\line(0,1){1}}\put(34,16){\line(0,1){1}}
\put(33,14){\line(0,1){1}}\put(34,14){\line(0,1){1}}
\put(33,3){\line(0,1){7}}\put(34,3){\line(0,1){7}}
\put(37,3){\line(0,1){10}}\put(38,3){\line(0,1){10}}
\put(28,13){\framebox(11,1)}
\put(32,15){\framebox(11,1)}
\put(41,16){\line(0,1){1}}\put(42,16){\line(0,1){1}}
\put(37,14){\line(0,1){1}}\put(38,14){\line(0,1){1}}
\multiput(23,5)(2,0){5}{\line(1,0){0.1}}
\multiput(17,11)(2,0){10}{\line(1,0){0.1}}
\multiput(17,19)(2,0){7}{\line(1,0){0.1}}
\put(37,16){\line(0,1){1}}\put(38,16){\line(0,1){1}}

\put(7,17){\framebox(36,1)}

\put(8,18){\line(0,1){1}}\put(9,18){\line(0,1){1}}
\put(41,18){\line(0,1){1}}\put(42,18){\line(0,1){1}}
\put(0,1){cell}\put(3.5,1){P}
\put(7.5,1){$-t$}\put(10,1){$-t+1$}\put(36.5,1){$t-1$}\put(41.5,1){$t$}
\put(0,5){$G_1$}\put(0,7){$G_1$} \put(0,13){$G_1$}\put(0,15){$G_1$}\put(0,17){$G_2$}

\put(2,5){\framebox(3,1)}\put(2,7){\framebox(3,1)}
\put(2,13){\framebox(3,1)}\put(2,15){\framebox(3,1)}
\put(5,5.5){\line(1,0){2}}\put(5,7.5){\line(1,0){6}}
\put(5,13.5){\line(1,0){23}}\put(5,15.5){\line(1,0){27}}

\put(3,3){\line(0,1){2}}\put(4,3){\line(0,1){2}}
\put(3,6){\line(0,1){1}}\put(4,6){\line(0,1){1}}
\put(3,8){\line(0,1){1}}\put(4,8){\line(0,1){1}}
\put(3,14){\line(0,1){1}}\put(4,14){\line(0,1){1}}
\put(3,16){\line(0,1){2}}\put(4,16){\line(0,1){2}}
\put(3,18){\line(0,1){1}}\put(4,18){\line(0,1){1}}
\end{picture}
\end{center}

Figure 1: The quantum circuit $C_2\circ C_1$ based on the set ${\cal G}=\{G_1,G_2\}$ of quantum gates. 
This circuit simulates one step of a QTM $M$ with amplitudes from ${\rm P}{\bf C}$. 

\end{figure}
Finally, we can exactly simulate any given finitely generated uniform QCF $\{C_n\}$ by a polynomial-time QTM $M$ with amplitudes from ${\rm P}{\bf C}$ (see Lemma 5.1 in \cite{NO02}) because the set of elementary gates for $C_n$ is finite and the quantum transition function of $M$ can represent all the operations induced by the elementary gates. 
Combining this fact with Theorem \ref{exact-equiv}, we obtain the perfect equivalence 
between the classes of polynomial-time QTMs with amplitudes from ${\rm P}{\bf C}$ 
and finitely generated uniform QCFs.  

\begin{Theorem}\label{perfect-equiv}
The class of polynomial-time QTMs with amplitudes from ${\rm P}{\bf C}$ 
is perfectly equivalent to the class of finitely generated uniform QCFs. 
\end{Theorem}

The following corollary directly comes from Theorem \ref{perfect-equiv}.  

\begin{Corollary}
The class of languages recognized with certainty (resp.\ with zero-error and bounded-error) 
by finitely generated QCFs coincide with the corresponding complexity class {\bf EQP} 
(resp.\ {\bf ZQP} and {\bf BQP}) for polynomial-time QTMs.  
\end{Corollary}

\section{Concluding Remarks}\label{Discussions}
We introduced a subclass of uniform QCFs, finitely generated uniform QCFs, and showed that 
this subclass is perfectly equivalent to the class of polynomial-time QTMs with amplitudes 
from polynomial-time computable numbers. 
Here, we shortly point out relationships among uniformity notions under other conditions. 


{\bf Complexity of coding:} 
In the theory of circuit complexity, a number of uniformity notions (say, \cite{Bor77,BCH86,Ruz81}) were proposed according to machines for constructing circuit families. 
We use polynomial-time DTMs as machines for computing the codes of QCFs. 
Although we could take polynomial-time QTMs with amplitudes from ${\rm P}{\bf C}$, instead of polynomial-time DTMs, the resulting class of uniform QCFs does not change. 
This is because such polynomial-time QTMs can be exactly simulated by QCFs whose coding functions are deterministically polynomial-time computable by using Yao's construction. 
Furthermore, we can restrict machines for computing the codes of QCFs to logspace DTMs without changing the resulting class of uniform QCFs since, in the proof of Theorem \ref{exact-equiv}, 
the code of $C$ is computed by an $O(\log n)$-space DTM. This means that the finitely generated ``logspace-uniform'' QCF is perfectly equivalent to the finitely generated ``P-uniform'' (and even ``EQP-uniform'') QCF, like the classical case. 


{\bf Parameter of uniform QCFs:} In \cite{KW00}, the binary string $x\in\{0,1\}^*$ was used 
as the parameter of uniform QCFs instead of the number $n\in{\bf N}$. 
That is, a QCF $\{C_x\}$ is uniform if there is a DTM $M$ that on input $x$, produces a description of $C_x$ in time polynomial in $|x|$.  However, as mentioned in \cite{KW00}, this variation does not change the resulting class of uniform QCFs. 
Actually, all the $C_x$'s with $|x|=n$ can be simulated by only one quantum circuit $C'_n$, 
which implements $C_x$ on the target part of $C'_n$ when $x$ is provided to the controlled part of $C'_n$. 
It is easy to check that $\{C'_n\}$ is a uniform QCF in the regular meaning. 


{\bf Size of elementary gates:} It is well-known that any one-qubit gates whose components are in ${\rm P}{\bf C}$ and the controlled-not gate are sufficient \cite{BBC$^+$95} to represent any uniform QCF exactly.   
Therefore, uniform QCFs defined under elementary gates of larger size more than two 
are perfectly equivalent to our uniform QCFs. 


{\bf Restriction of transition amplitudes:} 
In \cite{NO02}, we showed the perfect equivalence between semi-uniform QCFs and polynomial-time QTMs 
without any restriction of amplitudes. 
In Theorem \ref{exact-equiv}, we also have shown that both models are perfectly equivalent 
if their transition amplitudes are restricted to ${\rm P}{\bf C}$. 
However, if we restrict their transition amplitudes to other subsets of ${\bf C}$, 
there is no guarantee that two models are still perfectly equivalent. 
This is because, in the proof of Theorem \ref{exact-equiv}, we use the square root operation and the root of a function generated by trigonometric functions as well as the addition, the subtraction, the multiplication, and the division. 
By checking carefully the proof of Theorem \ref{exact-equiv}, we can establish the perfect equivalence 
between two models with amplitudes from the set of algebraic numbers, provided that the size of elementary gates is allowed to be any number fixed with respect to the input length. 
The case of the rational number field is open since Yao's construction needs the square root operation. 


In \cite{NO05}, the following related results on the power of finitely generated uniform QCFs 
have been reported by comparing it with that of uniform QCFs: 
(i) The quantum Fourier transform (QFT) of any order cannot be implemented with zero error 
by any finitely generated uniform QCF, while it can be exactly implemented by a uniform QCF 
by the result by Mosca and Zalka \cite{MZ03}. 
(ii) If a permutation $M_f$: $|x\rangle\mapsto|f(x)\rangle$ can be implemented with zero error by a uniform QCF, 
then both $f$ and $f^{-1}$ can be exactly computed by uniform QCFs.  
The first result implies that uniform QCFs cannot always be simulated with zero error 
by finitely generated uniform QCFs.\footnote{
Similarly, we can also show that uniform QCFs cannot always be simulated with zero error 
by finitely generated uniform QCFs with classical polynomial advice.} 
The second result suggests that uniform QCFs are more useful 
for constructing exact quantum algorithm than finitely generated QCFs. 

\

{\bf Acknowledgements:} 
M.O. gratefully acknowledges the financial support of the SCOPE project of the MIC,
the Grant-in-Aid for Scientific Research (B) 17340021 of the JSPS,
and the CREST project of the JST.


\begin{thebibliography}{99}
\setlength{\itemsep}{-1mm}
\bibitem{ADH97}
L. M. Adleman, J. DeMarrais, M. A. Huang, Quantum computability, SIAM J.\ Comput.\ {26} (1997) 1524--1540.

\bibitem{BBC$^+$95}
A. Barenco, C. H. Bennett, R. Cleve, D. P. DiVicenzo, N. Margolus, P. Shor, 
T. Sleator, J. Smolin, H. Weinfurter, Elementary gates for quantum computation, 
Phys.\ Rev. A {52} (1995) 3457--3467.

\bibitem{BCH86}
P. W. Beame, S. A. Cook, and H. J. Hoover, Log depth circuits for division and related problems, 
SIAM J.\ Comput.\ {15} (1986) 994--1003.

\ignore{\bibitem{Benn73}
C. H. Bennett, Logical reversibility of computation, 
IBM J.\ Res.\ Develop.\ {17} (1973) 525--532.}

\bibitem{BV93}
E. Bernstein, U. Vazirani, Quantum complexity theory (Preliminary abstract), 
In Proceedings of the 25th ACM Symposium on Theory of Computing, ACM Press, New York, 1993, pp.\ 11--20. 

\bibitem{BV97}
E. Bernstein, U. Vazirani, Quantum complexity theory, SIAM J.\ Comput.\ {26} (1997) 1411--1473. 
\ignore{
Preliminary version appeared in }

\bibitem{Bor77}
A. Borodin, On relating time and space to and size and depth, SIAM J.\ Comput.\ {6} (1977) 733--743.

\ignore{
\bibitem{BHMT00} G. Brassard, P. H{\o}yer, M. Mosca, A. Tapp, Quantum amplitude amplification and estimation, 
Quantum Computation and Quantum Information: A Millennium Volume. AMS Contemporary Mathematics Series, 
305 (2002) 53--74. Also {\tt quant-ph/0005055} (The quant-ph preprints are available at 
{\tt http://arxiv.org/abs/quant-ph/number}.)}

\bibitem{Deu85}
D. Deutsch, Quantum theory, the Church-Turing principle and the universal quantum computer, 
Proc.\ Roy.\ Soc.\ London Ser.\ A {400} (1985) 96--117.

\bibitem{Deu89}
D. Deutsch, Quantum computational networks,  
Proc.\ Roy.\ Soc.\ London Ser.\ A {425} (1989) 73--90.

\bibitem{EJ96}
A. Ekert, R. Jozsa, Shor's quantum algorithm for factoring numbers, 
Rev.\ Modern Phys.\ 68 (1996) 733--753. 

\ignore{
\bibitem{FR98} L. Fortnow, J. Rogers, Complexity limitations on quantum computation, 
J.\ of Comput.\ and System\ Sci.\ {59} (1999) 240--252.
}

\ignore{
\bibitem{Gro96} L. K. Grover, A fast quantum mechanical algorithm for database search,  
In Proceedings of the 28th ACM Symposium on Theory of Computing, ACM Press, New York, 1996, pp.\ 212--219.}

\bibitem{GHMP02}
F. Green, S. Homer, C. Moore, C. Pollett, Counting, fanout, and the complexity of quantum ACC, 
Quantum Information and Computation {2} (2002) 35--65.

\bibitem{Gru99}
J. Gruska, Quantum computing, McGraw-Hill, 1999. 

\ignore{
\bibitem{KKVB01} E. Kashefi, A. Kent, V. Vedral, K. Banaszek, Comparison of quantum oracles. Phys.\ Rev.\ A {65} 
(2002) 050304.
}

\bibitem{KF82}
Ker-I. Ko, H. Friedman, Computational complexity of real functions,  
Theoret.\ Comput.\ Sci.\ {20} (1982) 323--352.

\bibitem{Ko91}
Ker-I. Ko, Complexity Theory of Real Functions, Birkh\"{a}user, 1991.  

\bibitem{KW00}
A. Kitaev, J. Watrous, Parallelization, amplification, 
and exponential time simulation of quantum interactive proof systems,  
In Proceedings of the 32nd ACM Symposium on Theory of Computing, ACM Press, New York, 2000, pp.\ 608--617. 

\ignore{
\bibitem{Kit95} A. Kitaev, Quantum measurements and the abelian stabilizer problem, {\tt quant-ph/9511026}.}

\bibitem{MZ03} M. Mosca, C. Zalka, Exact quantum Fourier transforms and discrete logarithm algorithms, 
International Journal of Quantum Information {2} (2004) 91--100.

\bibitem{MV05} M. M\"{o}tt\"{o}nen, J. J. Vartiainen, Decomposition of general quantum gates, 
{\tt quant-ph/0504100}.

\ignore{
\bibitem{NC97} M. A. Nielsen, I. L. Chuang, Programmable quantum gate arrays, Phys.\ Rev.\ Lett.\ {79} (1997) 321--325. }
 
\bibitem{NC00} M. A. Nielsen, I. L. Chuang, Quantum Computation and Quantum Information, Cambridge, 2000.

\bibitem{Nis03} H. Nishimura, Quantum computation with restricted amplitudes, 
Int.\ J.\ Found.\ Comput.\ Sci.\ {14} (2003) 853--870.  

\bibitem{NO02}
H. Nishimura, M. Ozawa, Computational complexity of uniform quantum circuit families 
and quantum Turing machines, Theoret.\ Comput.\ Sci.\ {276} (2002) 147--181.

\bibitem{NO05}
H. Nishimura, M. Ozawa, Uniformity of quantum circuit families for error-free algorithms, 
Theoret.\ Comput.\ Sci.\ {392} (2005) 487--496.  

\bibitem{ON00} M. Ozawa, H. Nishimura, Local transition functions of quantum
Turing machines, RAIRO Theoret.\ Informatics Appl.\ {34} (2000) 379--402.

\bibitem{Pap94} C. H. Papadimitriou, Computational Complexity, Addison-Wesley, Reading, MA, 1994. 

\bibitem{Ruz81} W. L. Ruzzo, On uniform circuit complexity, J.\ Comput.\ Syst.\ Sci.\ {21} (1981) 365--383.  

\bibitem{Sho97}
P. W. Shor, Polynomial-time algorithms for prime factorization and discrete logarithms on a quantum computer,  
SIAM J. Comput. {26} (1997) 1484-1509. 
\ignore{
Preliminary version appeared in Proceedings of 35th IEEE Symposium on Foundations of Computer Science, 
IEEE Computer Society Press, Los Alamitos, CA, 1994, pp.\ 124--134. }

\bibitem{Sho96}
P. W. Shor, Fault-tolerant quantum computation, In Proceedings of the 37th IEEE Symposium 
on Foundations of Computer Science, IEEE Computer Society Press, Los Alamitos, CA, 1996, pp.\ 56--65. 

\ignore{
\bibitem{Tof80} T. Toffoli, Reversible computing, In Proceedings of the 7th International Colloquium 
on Automata, Languages and Programming, Lecture Notes in Comput.\ Sci.\ {84} (1980) 632--644.
}

\bibitem{Yam99}
T. Yamakami, A foundation of programming a multi-tape quantum Turing machine, In Proceedings 
of the 24th International Symposium on Mathematical Foundations of Computer Science,  
Lecture Notes in Comput.\ Sci.\ {1672} (1999) 430--441.

\bibitem{Yao93}
A. C.-C. Yao, Quantum circuit complexity, In Proceedings of the 34th Annual IEEE Symposium on 
Foundations of Computer Science, IEEE Computer Society Press, Los Alamitos, CA, 1993, pp.\ 352--361.
\end{thebibliography}
\end{document}